\documentclass[11pt]{article}
\usepackage{moriond,epsfig}
\usepackage[]{graphicx}
\usepackage[]{amssymb}

\def\be{\begin{equation}}
\def\ee{\end{equation}}
\def\bea{\begin{eqnarray}}
\def\eea{\end{eqnarray}}

\begin{document}
\vspace*{4cm}
\title{
Flavour puzzle or\\
Why neutrinos are different?}

\author{M.~Libanov$^{a}$ and F.-S.~Ling$^{b}$}

\address{
$^{a}$Institute for Nuclear Research of the Russian Academy of Sciences,\\
60th October Anniversary Prospect, 7a, 117312 Moscow, Russia  \\
$^{b}$Institute of Theoretical Physics,
Chinese University of Hong Kong
}

\maketitle\abstracts{
We present a short review of a 6-dimensional model where a flavour puzzle
of the Standard Model fermions finds an elegant solution. The mechanism is
based on an idea that the three fermionic generations originate from a
single 6D family. The model explains in a natural way both charged fermions
mass hierarchy and small mixings in the quark sector, and tiny neutrino
masses and large neutrino mixings. We also discuss some phenomenological
aspects of the model which can distinguish this class of the models from
another one and can help to look for manifestations of new physics at
colliders and other experiments.
}

\section{Introduction}
One of the most intriguing issues of the Standard Model (SM) is a flavour
puzzle which can be formulated  as the following three
problems:
\begin{itemize}
 \item {\em Problem of families replication and mass hierarchy:}
Why are there three families of fermions in the SM? In particular,
why are these generations differing only by masses and why is this
difference so large ($m_{\mathrm{top}}/m_{\mathrm{up}}\sim 10^{4}$)?
Why are mixings in the quark sector relatively small and why is the mixing
between first and third generations suppressed compare to the mixings
between adjacent generations?
\item {\em Neutrino mass problem:} Why do neutrinos have
tiny masses and large mixings? Why neutrinos are so different from charged
fermions?
\item {\em Flavour-changing neutral currents (FCNC) problem:} Why
we do not observe ``horizontal'' inter-generation transitions?
\end{itemize}

This paper is a short review of existing
works.\cite{Libanov:2000uf}$^{-}$\cite{Frere:2010ah} In these papers in
the frameworks of ''large extra dimensions'' (LED) (see
Ref.\cite{Rubakov:2001kp} for a review) an elegant solution to the flavour
puzzle has been suggested. The basic idea is an assumption that {\em
three} generations of SM fermions appear as three zero modes of {\em
single} multi-dimensional vector-like (with respect to SM gauge group)
family. In the current review we concentrate on main points and basic
ideas of the model at the cost of loss of mathematical rigor. For more
complete and more advanced details the Reader is directed to the original
works.

\section{The Setup}
\label{Section/Pg2/1:mor2011/The Setup}

Suppose one has single fermionic generation in a multi-dimensional theory.
Let us consider a topological defect whose core corresponds to our
four-dimensional world. Chiral fermionic zero modes may be trapped in the
core due to specific interaction with the fields which build up the
defect. In some cases, the index theorem guarantees that the number of
chiral zero modes is determined by the topological number of the defect
and by the charge of the fermion with respect to the symmetry group of
the fields forming the defect. We use this property to obtain three
fermionic generations localized on a defect while having only one
generation in the bulk. If the Brout-Englert-Higgs scalar couples to the
defect, it can also be trapped in the core. Hierarchy between masses of
three fermionic modes arises due to their different profiles in extra
dimensions.\cite{Libanov:2000uf} \cite{Frere:2000dc} \cite{Libanov:2002tm}

To be specific, let us assume that there are two large additional
dimensions.\footnote{The number of extra dimensions may be larger than
two. What we in fact assume is that the size of another additional
dimensions much smaller than the size of the two dimensions under
consideration.} The topological defect is an gauge
vortex. A principal issue of models with LED is the localization
of the SM gauge fields. One of possible ways to avoid this problem is to
consider the transverse extra dimensional space as a compact manifold and
to allow gauge fields to propagate freely in the extra dimensions. In what
follows we assume that the extra dimensions form a two dimensional sphere
with radius $R$.\cite{Frere:2003yv} Though gravity is not included in the
consideration, it should be stressed that the choice of the manifold is
not important for our principal conclusions.\cite{Frere:2003ye} The extra
dimensions can even be infinitely large. In this case, the role of the
radius $R$ of the sphere is taken by a typical size of the localized gauge
zero modes but {\em not} by a size of the extra dimensions.

\begin{table}[t]
\begin{center}
\begin{tabular}{|rc|c|c|c|c|c|}
\hline
\multicolumn{2}{|c|}{Fields}
& Profiles&\multicolumn{2}{|c|}{Charges}&
\multicolumn{2}{|c|}{Representations}\\
\cline{4-7}
&&&$U_g(1)$&$U_Y(1)$&$SU_W(2)$&$SU_C(3)$\\
\hline
 scalar&$\Phi$&$F(\theta ){\rm e}^{i\varphi }$&+1&0&{\bf 1}&{\bf
1}\\
&&$F(0)=0$, $F(\pi )\simeq v$&&&&\\
\hline
 vector&$A_{\varphi }$&$A(\theta )/e$&0&0&{\bf 0}&{\bf
0}\\
&&$A(0)=0$, $A(\pi )=1$&&&&\\
\hline
 scalar&$X$&$X(\theta )$&+1&0&{\bf 1}&{\bf 1}\\
&&$X(0)=v_X$, $X(\pi )=0$&&&&\\
\hline
 scalar&$H$&$H(\theta )$&--1&+1/2&{\bf 2}&{\bf 1}\\
&&$H_{i}(0)=\delta _{2i}v_H$, $H_{i}(\pi )=0$&&&&\\
\hline
 fermion&$Q$&3 L zero modes& axial $(3,0)$&$+1/6$&{\bf
2}&{\bf 3}\\
\hline
 fermion&$U$&3 R zero modes& axial $(0,3)$&$+2/3$&{\bf
1}&{\bf 3}\\
\hline
 fermion&$D$&3 R zero modes& axial $(0,3)$&$-1/3$&{\bf
1}&{\bf 3}\\
\hline
 fermion&$L$&3 L zero modes& axial $(3,0)$&$-1/2$&{\bf
2}&{\bf 1}\\
\hline
 fermion&$E$&3 R zero modes& axial $(0,3)$&$-1$&{\bf
1}&{\bf 1}\\
\hline
 fermion&$N$&{ Kaluza-Klein spectrum,}& {\ 0}&{ 0}&{\bf
1}&{\bf 1}\\
 &&{ no zero mode}& &&&\\
\hline
SM gauge&$\gamma ,G$&{ Kaluza-Klein spectrum}& {\ 0}&{ --}&{\bf
--}&{\bf --}\\
bosons &$Z,W^\pm$&{ starting from zero}& &&&\\
\hline
\end{tabular}
\end{center}
\caption{\label{Tab/Pg3/1:mor2011}Field content of the model.
For convenience, we describe here also the fields profiles in extra
dimensions. $\theta $ and $\varphi $ are the polar and the azimuthal
angles on the sphere, respectively. The vortex is localized
at $\theta =0$.
}
\end{table}

\begin{figure}
\centering
\includegraphics[width=0.45\textwidth,keepaspectratio,]{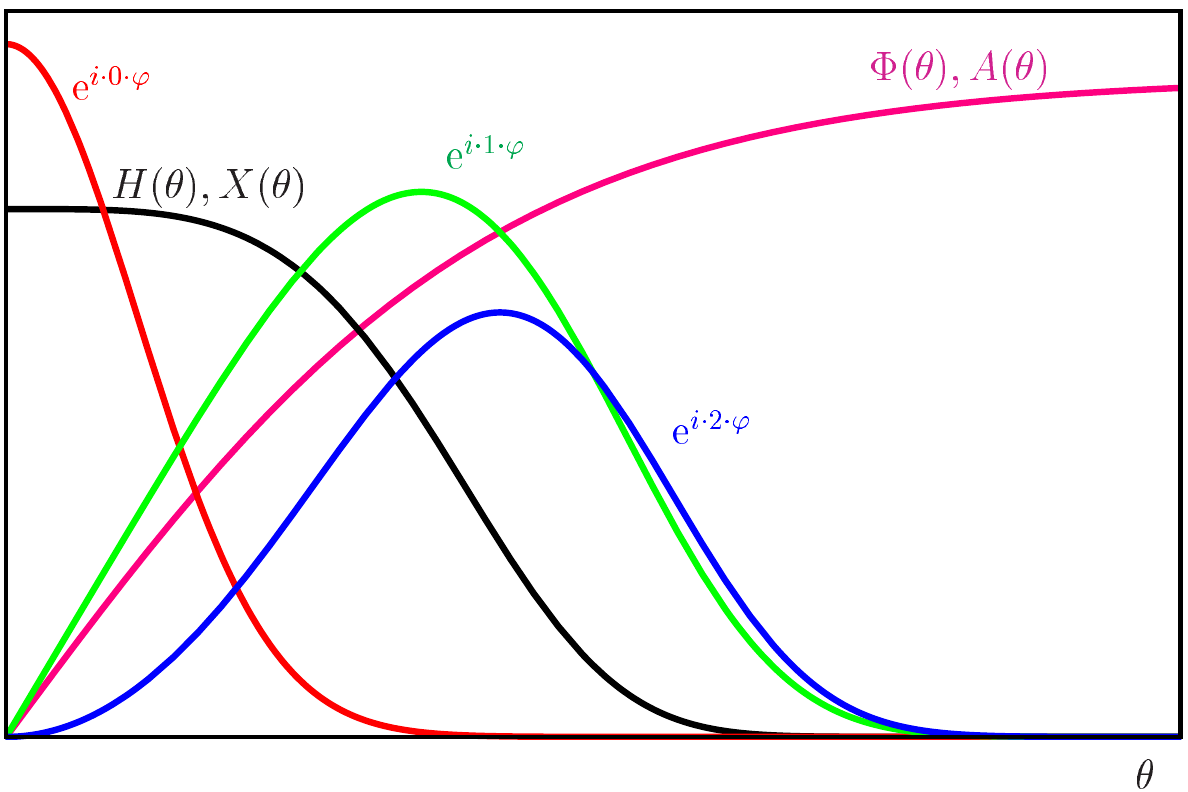}
\caption{Sketch of the relevant background fields profiles (black and
pink) and the fermionic zero modes (red, green and blue). The latter are
labeled by their $\varphi $-dependencies.
\label{Fig/Pg4/1:mor2011}
}
\end{figure}

The matter field content of the model is summarized in
Table~\ref{Tab/Pg3/1:mor2011} and the profiles of the relevant fields are
sketched in Fig~\ref{Fig/Pg4/1:mor2011}. The scalar field $\Phi $,
together with $U(1)_{g}$ gauge field, forms a vortex, while two other
scalars, $X$ and $H$, develop profiles localized on the vortex. There is
{\em one} fermionic generation which consists of five six-dimensional
fermions $Q$, $U$, $D$, $L$, and $E$. Each of the fermions develops, in the
vortex background, three chiral zero modes localized in the core of the
vortex, which correspond to three generations of the SM fermions. There is
an additional fermion $N$ which is neutral both under $U(1)_{g}$ and SM
gauge group. This fermion is not localized and its Kaluza-Klein (KK) modes
play a role of sterile neutrinos.

The spinor-scalar couplings, responsible for the localization of the
fermionic zero modes, are\footnote{We use the chiral representation for
the six-dimensional Dirac $\Gamma $-matrices (see
Ref.\cite{Libanov:2000uf} for notations). In particular, $\Gamma _{7}=
\mathrm{diag}(\mathbf{1},\mathbf{-1})$ is a six-dimensional analog of the
four-dimensional $\gamma _{5}$.}
\begin{eqnarray}
\sum \limits_{\Psi =Q,L}^{} g_{\Psi }\Phi ^{3}\bar{\Psi } \frac{1-\Gamma
_{7}}{2}\Psi +\mathrm{h.c.} &\longrightarrow& \mbox{3 left handed zero
modes}\nonumber\\
\sum \limits_{\Psi =U,D,E}^{} g_{\Psi }\Phi ^{3}\bar{\Psi } \frac{1+\Gamma
_{7}}{2}\Psi +\mathrm{h.c.} &\longrightarrow& \mbox{3 right handed zero
modes}
\label{Eqn/Pg4/1:mor2011}
\end{eqnarray}
This coupling may look surprising, as it is obviously non-renormalisable
in 4 dimensions. We are not considering renormalisability of the theory
here, as the 6-dimensional context is most likely an effective model, but
even in this case it might be suitable that the 4-dimensional reduction be
renormalisable.

What is important here is that three fermionic zero modes localized in the
vortex background have different generalized (supplemented by $U(1)_{g}$
global rotations) angular momentum\cite{Frere:2003yv}
and, as a result, have different $\varphi $ and $\theta $-dependencies.
Typically, one has
\begin{equation}
\Psi _{n}(\theta ,\varphi )\sim f_{n}(\theta
)\mathrm{e}^{i\left(3-n \right)\varphi }\;,\ \ \ \ n=1,2,3\;,
\label{Eq/Pg5/1:mor2011}
\end{equation}
with the $\theta $-dependent wave functions $f_{n}(\theta )$ behaving near
the origin as:
\begin{equation}
f_{n}(\theta )\sim \theta ^{3-n}\;,\ \ \ \ \theta \to 0\;.
\label{Eq/Pg5/2:mor2011}
\end{equation}

\section{Mass hierarchy and mixings}
\label{Section/Pg5/1:mor2011/Mass hierarchy and mixings}

The couplings (\ref{Eqn/Pg4/1:mor2011}) are responsible only for the
localization of the fermionic zero modes. From the 4-dimensional point of
view these modes describe localized {\em massless} chiral fermions with the
usual quantum numbers under SM gauge group.  Now comes the time to
generate the ``usual'' fermion masses, and to break the $SU(3)\times
SU(2)\times U(1)$ symmetry. This is done in the usual way, through the
Brout-Englert-Higgs mechanism, at the price of introducing a scalar
doublet, which we name $H$. In fact, for the purpose of separating the
various quantum numbers, we write (e.g., for the ``down'' quarks), instead
of the usual coupling:
\begin{equation}
\mathcal{ L}_{\mathrm{Yukawa}}=Y_{d}HX\bar{Q}\frac{1-\Gamma _{7}}{2}D
+\varepsilon _{d}Y_{d}H\Phi \bar{Q}\frac{1-\Gamma
_{7}}{2}D+\mathrm{h.c.}\;,
\label{Eq/Pg5/3:mor2011}
\end{equation}
where $Y_{d}$ and $\varepsilon _{d}Y_{d}$ are coupling constants, the
vortex scalar $\Phi $ has winding number 1 (see
Table~\ref{Tab/Pg3/1:mor2011}), while $X$ and $H$ have non-vanishing
values at $\theta =0$ and zero winding number.

The reduction to 4 dimensions involves integration over $\varphi $ and
$\theta $ and generates the mass terms. The first term in
(\ref{Eq/Pg5/3:mor2011}) yields
\begin{equation}
m_{D,nm}^{(1)}\propto\int \limits_{0}^{\pi }d\theta\sin\theta  \int
\limits_{0}^{2\pi }d\varphi f_{Q,n}^{\dag }f_{D,m}X(\theta )H(\theta
)\mathrm{e}^{i(n-m)\varphi }\;.
\label{Eq/Pg5/4:mor2011}
\end{equation}
Clearly, the integration over $\varphi $ guarantees that only diagonal
entries occur. The integral over $\theta $ is saturated near the origin,
more precisely, in the region where the fields $H$ and $X$ are non-zero.
This region coincides, as it is shown in Fig.~\ref{Fig/Pg4/1:mor2011}, with
the region $[0,\theta _{\Phi }]$ where $\Phi $ is appreciably different
from its VEV.\cite{Libanov:2002tm} \cite{Libanov:2007zz} In this region
one can safely use Eq.~(\ref{Eq/Pg5/2:mor2011}) for the fermionic wave
functions and finds\cite{Frere:2003yv}
\begin{equation}
m_{{D,nm}}^{(1)}\sim \delta _{nm}\sigma ^{2(3-n)}\;,
\label{Eq/Pg6/1:mor2011}
\end{equation}
where $\sigma =\theta _{\Phi }/\theta _{A}$ and $\theta _{A}\sim 1$ is
the typical size of the gauge field of the vortex.

Due to the non-trivial $\varphi $-dependence of $\Phi $ the second term in
(\ref{Eq/Pg5/3:mor2011}) results in non-diagonal elements of the mass
matrix:
\begin{equation}
m_{D,nm}^{(2)}\sim \varepsilon _{d}\delta _{n+1,m}\sigma ^{2(3-n)-1}
\label{Eq/Pg6/2:mor2011}
\end{equation}
The mass eigenstates are obtained by diagonalization
(\ref{Eq/Pg6/1:mor2011}), (\ref{Eq/Pg6/2:mor2011}) and the power-like
hierarchical mass pattern
\[
m_{33}:m_{22}:m_{11}=1:\sigma ^{2}:\sigma ^{4}
\]
arises at $\sigma \simeq 0.1$. The CKM-matrix has the form
\[
U^{\mbox{\small CKM}}\!\sim\!
\left(
\begin{array}{ccc}
1&\sigma&\sigma ^{4}\\
\sigma &1&\sigma \\
\sigma^{2} &\sigma &1
\end{array}
\right)
\]
and reproduces  observed mixings in the quark sector of the SM.

\section{Neutrinos masses. Why are they different?}
\label{Section/Pg6/1:mor2011/Neutrinos masses}

Now we want to consider whether the scheme can be extended to accommodate
mass and mixing data in the neutrino sector.
An obvious possibility to generate neutrino masses would be to treat them
exactly like the charged fermions, with a Dirac mass obtained at the
cost of introducing a 6D field $N$, bound to the vortex, and from which
the three families of 4D right-handed neutrinos emerge. However, this
possibility does not offer a natural explanation for the smallness of the
neutrino masses, which in this case require tiny coefficients in the
Lagrangian.

It is therefore tempting to consider other solutions, namely the case
where the ``right hande'' neutrino field is {\em NOT} bound to the vortex.
In the context of models with LED, tiny neutrino masses
are often the result of a dilution effect: the field that provides
right hande neutrinos, being singlet under the SM gauge
group, can be non-localized, and therefore have a small overlap with the
wave function of SM fields. A first attempt using a non-localized field $N$
was made in Ref.\cite{Frere:2001ug} However, that approach predicts a
neutrino mixing pattern that differs significantly from the observed one.

Here we consider another possibility.\cite{Frere:2010ah} We assume that
spinor $N$ is a gauge singlet both under SM and $U(1)_{g}$ gauge group. It
means first of all, that it can freely propagate in the extra dimensions.
Secondly, one can (and, in general, have to) write a Majorana-like mass
term in 6D for it:
\[
\frac{M}{2}(\bar{N}^{c}N +\bar{N}N^{c}).
\]
From 4-dimensional point of view one has KK tower of fermions possessing
4D Majorana mass $M$ and different Dirac masses starting from
$1/R$.\footnote{It is worth noting that a possible 6D Dirac mass term does
not play any role and translates to a shift of 4D Dirac mass spectrum.
This is the reason why we do not consider it.}

Thirdly, the gauge invariance allows one to introduce the following
couplings ($\tilde{H}=i\sigma _{2}H^{*}$)
\[
\sum \limits_{S_{+} }^{}Y^+_{\nu,S_+}
\tilde{H} S_+ \bar{L} {1 + \Gamma_7 \over 2} N + \sum
\limits_{S_{-}}^{}Y^-_{\nu,S_-} \tilde{H} S_- \bar{L} {1 - \Gamma_7 \over
2} N + {\rm h.c.} \quad ,
\]
where $S_+$ and $S_-$ have $U(1)_{g}$ gauge charges 1 and $-2$,
respectively, and can be\footnote{In the quark sector
(Sec.~\ref{Section/Pg5/1:mor2011/Mass hierarchy and mixings}) we
restricted ourselves to considering $S_{+}=\Phi ^{*},X^{*}$ only. The
reason is that an inclusion of more composite structures (e.g., $S_{-}$)
does not play any significant role.}
\begin{eqnarray}
S_+ &=& X^*, \; \Phi^*, \; X^{*2} \Phi, \; \dots \nonumber \\
S_- &=& X^2, \; X \Phi, \; \Phi^2, \; \dots \label{Eqn/Pg7/1:mor2011}
\end{eqnarray}
In 4D these couplings give rise to mixings between heavy modes of $N$ and
zero modes of active SM neutrinos. Together with the Majorana mass of
modes of $N$ it winds up a ``see-saw'' mechanism yielding tiny
Majorana masses of the active neutrinos. The resulting neutrino mass
matrix can be schematically written in the form:
\begin{equation}
m^{\nu }_{mn}\sim\int \limits_{0}^{\pi }d\theta \int \limits_{0}^{2\pi
}d\varphi F(\theta ,\varphi )\bar{L}_{n}^{c}L_{m}\;,
\label{Eq/Pg8/1:mor2011}
\end{equation}
where $F(\theta ,\varphi )$ is determined by $S_{\pm}$ as well as by wave
functions of $N$. The main point and the main difference from  the quark
sector (see Eq.~(\ref{Eq/Pg5/4:mor2011})) is the presence {\em charge
conjugated} spinor in the integrand: $\bar{L}^{c}\sim L^{T}$. This leads
to completely different from (\ref{Eq/Pg6/1:mor2011}),
(\ref{Eq/Pg6/2:mor2011}) selection rules. For instance, if we restricted
ourselves by $\varphi $-independent $S_{\pm}$ (the first structures in
(\ref{Eqn/Pg7/1:mor2011})) then $F$ does not depend on $\varphi $, and one
has
\begin{equation}
m^{\nu }_{mn}\sim \int \limits_{0}^{2\pi }d\varphi
\mathrm{e}^{i(4-n-m)\varphi }\sim \delta _{n,4-m}\sim
\left(
\begin{array}{ccc}
{ \cdot}& { \cdot}& 1\\
{ \cdot}& \sigma ^{2}&{ \cdot}\\
1&{ \cdot}&{ \cdot}\nonumber
\end{array}
   \right)
\label{Eq/Pg8/2:mor2011}
\end{equation}
The inclusion of the $\varphi $-dependent structures in
(\ref{Eqn/Pg7/1:mor2011}) gives rise to non-zero off secondary diagonal
elements which have at least an order of $\sigma $.

What are consequences  of the mass pattern (\ref{Eq/Pg8/2:mor2011})? The
neutrino mass matrix (\ref{Eq/Pg8/2:mor2011}) is diagonalized by a matrix
with the structure
\[
U_{\nu }
\sim \left(
\begin{array}{ccc}
1/\sqrt{2}&1/\sqrt{2} &\sigma \\
\sigma  &\sigma &1 \\
-1/\sqrt{2}&1/\sqrt{2}&\sigma
\end{array}
   \right)   +\mathcal{ O}(\sigma ^{2})\;.
\]
Let us emphasize that the large mixing angle in the 1--3 block is maximal
up to $\sigma ^{2}$ corrections. When the charged lepton mass matrix
contains a large mixing angle in the 2--3 block, this model predicts two
large mixing angles, as observed. The remaining small
mixing angle $U_{e3}$, which corresponds to the weight of the lightest mass
eigenstate in the electronic neutrino, is predicted to be of order $\sigma
\sim 0.1$.

The diagonalized neutrino mass matrix has the inverted hierarchy pattern
$\mathrm{diag}(m+\mathcal{ O}(\sigma ^{2}),-m + \mathcal{ O}(\sigma
^{2}),m\sigma ^{2})$. Therefore, this model naturally predicts a hierarchy
in the mass squared splittings relevant in neutrino oscillation
experiments $\Delta m_{12}^{2}/\Delta m_{13}^{2}\sim \sigma ^{2}\sim 0.01$,
in good agreement with the observed data $\Delta m_{12}^{2}/\Delta
m_{13}^{2}\simeq 3.2\% $. Moreover, $m_{1}$
and
$m_{2}$ form a ``pseudo-Dirac'' pair as $m_{1}+m_{2}\sim \sigma ^{2} m$.
It leads to a partial cancellation in the effective Majorana mass,
$|\langle m_{\beta \beta }\rangle| =|\sum
\limits_{i}^{}m_{i}U_{ei}^{2}|\simeq 1/3\sqrt{\Delta m_{13}^{2}}$,
defining the amplitude for neutrinoless double-beta decay.

\section{FCNC}
\label{Section/Pg9/1:mor2011/FCNC}

From the 4D point of view the presented model completely reproduces all
properties of SM if one considers zero modes only (including zero modes
of the SM gauge fields). In particular,  all FCNC processes are strongly
suppressed as it occurs in SM. However, from the 6D point of view we have
only a single generation and there is no difference, say, between $\mu $
and $e$. That is, heavy (non-zero) modes of the neutral SM gauge bosons
can (and have to) violate flavour and/or lepton numbers. Nevertheless,
without account of inter-generation mixings, the generalized angular
momentum or, what is the same, the generation number $G$ is exactly
conserved. This forbids all processes with nonzero change of $G$;  the
probabilities of the latters in the full theory are thus suppressed by
powers of the mass-matrix mixing parameter, $(\varepsilon \sigma)^{2\Delta
G} $. However, the amplitudes of processes with $\Delta G=0$ but lepton and
quark flavours violated separately are suppressed only by the mass squared
of the KK modes of the SM gauge bosons.

In Ref.\cite{Frere:2003ye} the following specific flavour violating
processes have been studied:
\begin{itemize}
 \item $\Delta G=0$: $K^{0\to }_{L}\to \mu e$, $K^{+}\to \pi
^{+}e^{-}\mu ^{+}$;
 \item $\Delta G=1$: $\mu \to e\gamma $, $\mu \to 3e$, $\mu \to
e$-conversion;
\item $\Delta G=2$: $K_{L}-K_{S}$ mass difference and CP violation in
kaons.
\end{itemize}
These processes are known to give the strongest constraints on masses and
couplings of new vector bosons. It was found that indeed the pattern of
flavour violation is distinctive: contrary to other models, processes with
change of the generation number $G$ by one or two units are strongly
suppressed compared to other rare processes.  The strongest constraint on
the model arises from non-observation of the decay $K^{0}\to \mu e$; it
requires that the size of the  sphere (size of the gauge-boson
localization) $R$ satisfies $ {\varkappa}/{R} \gtrsim 64 \mbox{ TeV,} $
where $\varkappa$ is a dimensionless parameter depending on specific
model: details of the geometry, mechanism of the localization of the
vector fields, and so on. A clear signature of the model would be an
observation of $K^{0}\to \mu e$ decay without observation of $\mu \to 3e$,
 $\mu \to e\gamma $ and $\mu e$-conversion at the same precision level.

For the spherical model under discussion $\varkappa=1$. However, in
general one can expect that $\varkappa$ can differ from unit and can be
small enough ($\varkappa\sim 0.01$). In the latter case the masses of the
first non-zero excitations of the SM gauge bosons $M'\simeq 1/R$ can be of
order of few TeV and vector bosons can, in principle, be observed at
colliders. In general, there are two possibilities. First of all, one can
try to search for ``usual'' heavy vector bosons, that is the heavy KK modes
which do not change generation number. The second possibility is to look
for heavy KK vector bosons due to the flavour-changing decay modes into
($\mu e$) or $(\mu \tau )$ pairs. The flavour-changing decays of this kind
have a distinctive signature: antimuon and electron (or their
antiparticles) with equal and large transverse momenta in the final state.

The latter possibility has been investigated in Ref.\cite{Frere:2004yu} In
particular, it was found for the expected LHC value of 100fb$^{-1}$
for luminosity and $\sqrt{s}=14$TeV that the number of $pp\to \mu
^{+}e^{-}$ events varies from 1 to 10 per year for $M'\simeq 3\div 1$TeV.
The probability of the production ($\mu ^{-}e^{+}$) pairs is approximately
ten times smaller due to the former process can use valence $u$ and
$d$-quarks in the proton, while the second only involves partons from the
sea. The same numbers are representative also for the ($\mu \tau $)
channels.

There are also other signatures of FCNC effects, in particular, with
hadronic final states, when ($\bar{t},c$) or ($\bar{b},s$) jets are
produced. The dominant contribution to these processes arises from the
interactions with higher KK modes of gluons, which have large coupling
constant. For the mass of $M' = 1$TeV the number of events has been
estimated as $N = 1.2 \cdot 10^{3}$. But potentially large SM backgrounds
should be carefully considered for such channels.

\section{Conclusions}

To conclude, we presented a possible elegant solution to the flavour
puzzle. The mechanism is based on an assumptions that the {\em three} SM
generations originate from a {\em single} family in a higher-dimensional
theory. The generation number is none other than angular momentum and,
therefore, has an geometrical origin.

We explained why neutrinos are different from the sector of the charged
fermions. A light neutrino mass matrix where one mixing angle is
automatically maximal and where the eigenvalues obey an inverted hierarchy
with a pseudo-Dirac pattern for the heavier states $m_{1}\simeq -m_{2}\gg
m_{3}$ is a result of a mixing between active neutrinos and a single heavy
sterile 6D fermion with {\em Majorana-like} mass.

The 6D Lagrangian with one generation contains much less parameters than
the effective one. All masses and mixings of the SM fermions are governed
by a few parameters of order one. This fact allows for specific
phenomenological predictions. In particular, the KK modes of the vector
bosons mediate flavour-violating processes. The pattern of flavour
violation is distinctive: contrary to other models, processes with change
of the generation number by one or two units are strongly suppressed
compared to other rare processes.  The strongest constraint on the model
arises from non-observation of the decay $K\to \mu e$; it requires that
the size of the extra-dimensional sphere (size of the gauge-boson
localization) satisfies $1/R \gtrsim 64$ TeV. The KK modes of vector
bosons have larger masses, but for large enough $R$, could be detected by
precision measurements at colliders.

One more point which we did not discuss in this review is a
Brout-Englert-Higgs boson properties. The model predicts that its mass can
not be much larger then 100 GeV.\cite{Libanov:2007zz}

\section*{Acknowledgments}
Many thanks to the conference organizers for the invitation.
We wish to thank very warmly our colleagues J.-M.~Frere, E.~Nugaev and
S.~Troitsky for our ongoing collaboration.  The work of M.L. has been
supported in part by the Federal Agency for Science and Innovations under
state contract 02.740.11.0244, by the grant of the President of the
Russian Federation NS-5525.2010.2, by the RFBR grants 11-02-08018 and
11-02-92108, and by the Dynasty Foundation.


\section*{References}

\begin{thebibliography}{13}

\bibitem{Libanov:2000uf}
  M.~V.~Libanov and S.~V.~Troitsky,
  Nucl.\ Phys.\  B {\bf 599}, 319 (2001).


\bibitem{Frere:2000dc}
  J.~M.~Frere, M.~V.~Libanov and S.~V.~Troitsky,
  Phys.\ Lett.\  B {\bf 512}, 169 (2001).

\bibitem{Libanov:2002tm}
  M.~V.~Libanov and E.~Y.~Nougaev,
  JHEP {\bf 0204}, 055 (2002); Surveys High Energ.\ Phys.\  {\bf 17}
(2002) 165.

\bibitem{Frere:2003yv}
  J.~M.~Frere, M.~V.~Libanov, E.~Y.~Nugaev and S.~V.~Troitsky,
  JHEP {\bf 0306}, 009 (2003).

\bibitem{Frere:2003ye}
  J.~M.~Frere, M.~V.~Libanov, E.~Y.~Nugaev and S.~V.~Troitsky,
  JHEP {\bf 0403}, 001 (2004).

\bibitem{Libanov:2007zz}
  M.~V.~Libanov and E.~Y.~Nugaev,
  Phys.\ Atom.\ Nucl.\  {\bf 70}, 864 (2007)
  [Yad.\ Fiz.\  {\bf 70}, 898 (2007)] [arXiv:hep-ph/0512223].

\bibitem{Frere:2001ug}
  J.~M.~Frere, M.~V.~Libanov and S.~V.~Troitsky,
  JHEP {\bf 0111}, 025 (2001).

\bibitem{Frere:2004yu}
  J.~M.~Frere, M.~V.~Libanov, E.~Y.~Nugaev and S.~V.~Troitsky,
  JETP Lett.\  {\bf 79}, 598 (2004)
  [Pisma Zh.\ Eksp.\ Teor.\ Fiz.\  {\bf 79}, 734 (2004)]
  [arXiv:hep-ph/0404139].

\bibitem{Frere:2010ah}
  J.~M.~Frere, M.~Libanov and F.~S.~Ling,
  JHEP {\bf 1009}, 081 (2010).

\bibitem{Rubakov:2001kp}
  V.~A.~Rubakov,
  Phys.\ Usp.\  {\bf 44}, 871 (2001)
  [Usp.\ Fiz.\ Nauk {\bf 171}, 913 (2001)]
  [arXiv:hep-ph/0104152].



\end{thebibliography}
\end{document}